# THEORETICAL STUDY OF COUPLING MECHANISMS BETWEEN OXYGEN DIFFUSION, CHEMICAL REACTION, MECHANICAL STRESSES IN A SOLID-GAS REACTIVE SYSTEM.

Nicolas CRETON, Virgil OPTASANU, Tony MONTESIN, Sébastien GARRUCHET

Institut Carnot de Bourgogne, UMR 5209 CNRS, Université de Bourgogne, 9 avenue Alain Savary, BP 47870, 21078 Dijon cedex France

**Abstract** - This paper offers a study of oxygen dissolution into a solid, and its consequences on the mechanical behaviour of the material. In fact, mechanical strains strongly influence the oxidation processes and may be, in some materials, responsible for cracking. To realize this study, mechanical considerations are introduced into the classical diffusion laws. Simulations were made for the particular case of uranium dioxide, which undergoes the chemical fragmentation. According to our simulations, the hypothesis of a compression stress field into the oxidised $UO_2$ compound near the internal interface is consistent with some oxidation mechanisms of oxidation experimentally observed. More generally, this work will be extended to the simulation to an oxide layer growth on a metallic substrate.

**Résumé – Etude théorique des mécanisme de couplage entre diffusion d'oxygène, reaction chimique et contraintes mécaniques dans un système réactif solide/gaz.** Le but de ce travail est d'introduire un modèle thermodynamique afin de décrire la croissance d'une couche d'oxyde à la surface d'un substrat métallique. Plus précisément, nous proposons ici une étude de la dissolution d'oxygène dans un solide, et ses conséquences sur l'apparition de contraintes mécaniques. Celles-ci influencent fortement les processus d'oxydation et sont responsables, dans certains matériaux, de l'apparition de fissures. Pour réaliser cette étude, des considérations mécaniques sont introduites dans les lois classiques de la diffusion. Des simulations ont été faites dans le cas particulier de l'oxyde d'uranium, qui subit la fragmentation chimique. Grâce à ces simulations, l'hypothèse d'un champ de contrainte de compression dans le composé oxyde d'$UO_2$, à proximité de l'interface, est cohérente avec l'interprétation des mécanismes d'oxydation observés expérimentalement.

## 1. INTRODUCTION

In contact with oxygen, some crystalline solids undergo a chemical transformation during which the cracking and fragmentation of the initial solid is observed. Known as "chemical fragmentation", this oxidation reaction induces mechanical strains due to strong interactions between the different mechanisms that occur during the oxide layer growth. They can be induced by the diffusion of species or by the chemical oxidation reaction. But they influence themselves the diffusion and the reaction, modifying particles flux and interfaces displacement velocities, governed by the thermodynamic forces (chemical potential gradients and affinity), which characterize these processes. In the way of the theoretical approach of such a process, a purely chemical formulation cannot explain by itself a phenomenon like the chemical fragmentation.

To solve such a problem, a mechano-chemical approach based on non-equilibrium thermodynamics has been developed [1]. Initially written for the study of Zr anionic oxidation, this model allows us to set a new expression for the equation governing the diffusion of species into the solid, and a new formulation of the substrate/oxide interface displacement including mechanical terms.

In the first part of this paper, thermodynamic bases of our model are reminded. Furthermore, this first part accents on the volume component of the intrinsic dissipation, which leads to the formulation of a new expression for the matter transport law. This one is then applied to a grain submitted to different strain fields in order to study its mechano-chemical behaviour. Some numerical simulations are realized to validate our equations; they allow us to verify the influence of the mechanics on the classical diffusion laws. This theoretical approach is the first step of a more global study. It will finally lead to the simulation of the different oxide phases growths at the surface of any material in which the chemical fragmentation occurs. In such a perspective, the values used in the simulations concern a particular oxide $UO_2$, which undergoes this phenomenon.

## 2. THERMODYNAMICS OF THE DIFFUSION-REACTION PROCESS

The construction of an original predictive model concerning an anionic oxidation process needs to obtain the evolution law governing the motion of the substrate/oxide interface, including mechanics and mass transport. For this, non-equilibrium thermodynamics is used to determine the intrinsic dissipation associated both with the diffusion of species into the metal and the motion of the reactive interface [2].

### 2.1. Dissipation of the system

Let's consider a quasi-static and isothermal evolution of an opened thermodynamic system V (*figure 1*) including a moving internal interface $\Sigma$. If $\Phi = E - TS$ is the Helmholtz free energy of the system, the first and second principles leads to the following dissipation equation:

$$\mathcal{D} = P_{ext} - \dot{\Phi} \geq 0 \qquad (1)$$

In this expression, $P_{ext}$ is the power developed by the forces emerging from the mass fluxes and the external strengths applied to the external boundary of the system. E is the internal energy, T the temperature and S the entropy. The intrinsic dissipation will allow us to obtain the evolution equations of the internal variables governing the oxidation transformation of the substrate. It is then necessary to find an expression for $\Phi$ and $P_{ext}$ introducing the variables and constants of the system.

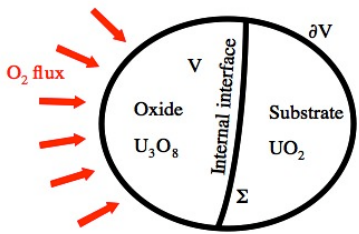

**Figure 1.** Schematic representation of the system.

### 2.2. Helmholtz free energy of the system

The free energy $\Phi$ is a function of three independent internal variables, strongly coupled:
- the elastic strain $\varepsilon_{ij}^e$,
- the concentration $c$,
- the reaction degree of conversion $\xi$.

It appears judicious to divide the expression of this energy in two parts, the one in volume to take into account the mechanical properties and diffusion in the system, and the other one in surface to represent the reaction at the interface $\Sigma$:

$$\Phi = \int_V \Phi_V(\varepsilon_{ij}^e, c) dV + \int_\Sigma \Phi_\Sigma(\xi) d\Sigma \qquad (2)$$

By introducing the Cauchy stress tensor $\sigma_{ij}$, the chemical potential $\mu_\gamma$ of a species $\gamma$ and the chemical affinity A of the reaction as:

$$\sigma_{ij} = \left(\frac{\partial \Phi_V}{\partial \varepsilon_{ij}^e}\right)_c, \mu_\gamma = \left(\frac{\partial \Phi_V}{\partial c}\right)_{\varepsilon_{ij}^e, c_{\beta \neq \gamma}}, A = \left(\frac{\partial \Phi_V}{\partial \xi}\right) \qquad (3)$$

we obtain then a first expression for the Helmholtz free energy [4]:

$$\Phi_V = \int \sigma_{ij} d\varepsilon_{ij}^e + \sum_\gamma \int \mu_\gamma dc_\gamma \text{, and } \Phi_\Sigma = \int A d\xi \qquad (4)$$

In $\Phi_V$, the first term corresponds to a mechanical contribution representing the differential of elastic energy density. The second term corresponds to a chemical contribution linked to the flux of mass diffusing in the system. $\Phi_\Sigma$ is a chemical contribution due to the reactions inside the system.

2.3. Helmholtz free energy evolution

The evolution of the Helmholtz free energy is due to the diffusion of some species inside the material, whose consequence is to gradually modify the substrate composition. When the limits in concentration of species are reached, the chemical reaction can occur, the oxide grows, and the boundary moves. An expression of $\dot{\Phi}$ is given below:

$$\dot{\Phi} = \int_V \left(\sigma_{ij} \dot{\varepsilon}_{ij}^e + \sum_\gamma \mu_\gamma \dot{c}_\gamma\right) dV - \int_\Sigma \left\{\left(\langle\sigma_{ij}\rangle[\varepsilon_{ij}^e] + \sum_\gamma \int_{c_{\gamma-}}^{c_{\gamma+}} \mu_\gamma dc_\gamma\right) \vec{\omega}\vec{n} + A\dot{\xi}\right\} d\Sigma \qquad (5)$$

where $\vec{\omega}$ represents the speed of propagation of the surface $\Sigma$ and $\vec{n}$ the normal to this surface. $\langle\sigma_{ij}\rangle = \frac{1}{2}(\sigma_{ij}^{ox} + \sigma_{ij}^{met})$ represents the average stress and $[\varepsilon_{ij}^e] = \varepsilon_{ij}^{e.ox} - \varepsilon_{ij}^{e.sub}$ the strain jump across the interface $\Sigma$: $\varepsilon_{ij}^{e.ox}$ and $\varepsilon_{ij}^{e.sub}$ are the limits of $\varepsilon_{ij}^e$ when the interface $\Sigma$ is approached from the oxide and from the substrate respectively.

2.4. Power developed by external forces

When volume forces are neglected, the oxygen flow through the external surface creates an external force acting on the surface and responsible for the occurence of an external stress ($\sigma_{ij}$) applied to the volume boundary. In other words, the power of external forces (and exclusively surface forces here) is due to:
- the forces applied to $\partial V$.
- the mass flux entering the system through this surface.

This power can be written as [1]:

$$P_{ext} = \int_V \left\{\sigma_{ij}\dot{\varepsilon}_{ij} - \sum_\gamma \left(\mu_\gamma div(\vec{J_\gamma}) + \vec{J_\gamma}.\vec{\nabla}\mu_\gamma\right)\right\} dV - \int_\Sigma \left\{\sigma_{ij}[\varepsilon_{ij}]\vec{\omega}\vec{n} + \sum_\gamma [\mu_\gamma \vec{J_\gamma} \vec{n}]\right\} d\Sigma \qquad (9)$$

2.5. Dissipation

Let's classically break up the strain into an elastic part and an inelastic part: $\varepsilon_{ij} = \varepsilon_{ij}^e + \varepsilon_{ij}^{inel}$ ($\varepsilon_{ij}^{inel}$ represents all the inelastic strain appearing during the oxidation of metal). Furthermore, the mass conservation relation can be written as $\dot{c}_\gamma = -div(\vec{J_\gamma})$. From equation (7) and equation (9), we can determine the dissipation expression of the system:

$$\mathcal{D} = \int_V \left\{ \sigma_{ij} \dot{\varepsilon}_{ij}^{inel} - \sum_\gamma \vec{J}_\gamma . \vec{\nabla} \mu_\gamma \right\} dV$$
$$- \int_\Sigma \left\{ \langle \sigma_{ij} \rangle [\varepsilon_{ij}^{inel}] \vec{\omega}.\vec{n} + \sum_\gamma [\mu_\gamma \vec{J}_\gamma].\vec{n} - \left( \sum_\gamma \int_{c_{\gamma-}}^{c_{\gamma+}} \mu_\gamma dc_\gamma \right) \vec{\omega}.\vec{n} + A\dot{\xi} \right\} d\Sigma \quad (10)$$

2.6. <u>Volume part of the dissipation</u>

The first integral in equation (10) corresponds to the volume part of the dissipation. It represents the energy dissipated in the volume, due to the oxygen diffusion process and the plastic strains generated inside the volume by the oxidation process. This integral should contain an evolution law similar to the Fick's law (in accordance with the hypothesis that diffusion takes place in volume). Let's consider the inelastic strain as $\varepsilon_{ij}^{inel} = \varepsilon_{ij}^{ch} + \varepsilon_{ij}^{p}$, where $\varepsilon_{ij}^{ch}$ corresponds to the deformation of the lattice due to diffusion of species and $\varepsilon_{ij}^{p}$ corresponds to every other kinds of deformation (plastic part). A definition of the chemical strain was given by Larché [5], who introduced a chemical expansion coefficient $\eta_{ij}^\gamma$ representing the deformation generated by the species γ diffusing in the material, per unit of concentration:

$$\eta_{ij}^\gamma = \left( \frac{\partial \varepsilon_{ij}^{ch}}{\partial c_\gamma} \right)_{\sigma_{ij}, T} \quad (11)$$

For example, in the particular case of $UO_2$, the coefficient $\eta_{ij}^{ox}$ is negative: the volume of $UO_2$ decreases when oxygen diffuses in it. From equation (10) and equation (11), and if $\eta_{ij}^\gamma$ does not depend on time and concentration, the volume dissipation becomes:

$$\mathcal{D}_V = \int_V \left\{ \sigma_{ij} \dot{\varepsilon}_{ij}^p + \sigma_{ij} \eta_{ij}^\gamma \dot{c}_\gamma - \sum_\gamma \vec{J}_\gamma . \vec{\nabla} \mu_\gamma \right\} dV \quad (12)$$

The two last terms of equation (12) correspond to the diffusion process inside the material: they must be governed by the same internal variables $\vec{J}_\gamma$. Considering an infinitesimal volume δV and introducing the definition of the flux $\vec{J}_\gamma = c_\gamma \vec{v}_\gamma$ ($\vec{v}_\gamma$ is the velocity of the species γ), we obtain:

$$\mathcal{D}_{\delta V} = \left\{ \sigma_{ij} \dot{\varepsilon}_{ij}^p - \sum_\gamma \vec{J}_\gamma . \left\{ \nabla(\mu_\gamma + \sigma_{ij} \eta_{ij}^\gamma) - \left( \frac{\partial \sigma_{ij} \eta_{ij}^\gamma}{\partial c_\gamma} + \frac{\sigma_{ij} \eta_{ij}^\gamma}{c_\gamma} \right) . \vec{\nabla} c_\gamma \right\} \right\} \quad (13)$$

In what follows, we consider the plastic strains are negligible: this is an acceptable hypothesis for a material showing an elastic behaviour as $UO_2$. Furthermore, noting that $\vec{\nabla}\mu_\gamma = (\partial \mu_\gamma / \partial c_\gamma) \vec{\nabla} c_\gamma$, and assuming the chemical potential definition given by Larché and Cahn in equation (14) [5], we obtain a final expression for the dissipation (equation 15) from which a flux law will ensue.

$$\mu_\gamma(\sigma_{ij}, T, c_\gamma) = \mu_\gamma^c(T, c_\gamma) - \sigma_{ij} \eta_{ij}^\gamma \quad (14)$$

$$\mathcal{D}_{\delta V} = -\vec{J}_1 . \left\{ \left( \frac{\partial \mu_1^c}{\partial c_1} \right) - \left( \frac{\partial \sigma_{ij} \eta_{ij}^1}{\partial c_1} + \frac{\sigma_{ij} \eta_{ij}^1}{c_1} \right) \right\} . \vec{\nabla} c_1 \quad (15)$$

In equation (14), $\mu_\gamma^c(T,c_\gamma)$ represents the chemical potential only depending on concentration and temperature. Equation (15) corresponds to diffusion of only one species (oxygen). The thermodynamic force driving the diffusion process is decomposed into a purely chemical part and a mechano-chemical part. From this equation, it is possible to obtain an evolution law looking like a Fick's law. So, let's now consider that:
- the fluxes are linear function of the forces (Onsager near –equilibrium conditions),
- the flux can be written in a classical way as $\vec{J_1} = -D_1' \vec{\nabla} c_1$,
- the definition of the pure chemical potential is $\mu_1^c = \mu_1^0 + RT \ln(x_1)$, it comes:

$$D_1' = D_1 \left\{ 1 - \frac{\eta_{ij}^1 c_1}{RT} \left( \frac{\partial \sigma_{ij}}{\partial c_1} + \frac{\sigma_{ij}}{c_1} \right) \right\} \quad (16)$$

The diffusion coefficient $D_1'$ can be separated into two components:
- The first one corresponds to the classical diffusion coefficient in a stress-free state: $D_1' = D_1$. In this case, equation (16) corresponds to a classical Fick's law.
- The second one depends on both the stress state and the composition. This Nernst term conveys the forced diffusion induced by stresses during the diffusion of species inside the material. It directly influences the diffusion coefficient: according to the stress and/or its evolution, the oxygen dissolution will be speeded-up or slowed down.

3. NUMERICAL SIMULATIONS

To validate our hypothesis, we will apply numerical evaluation of our model to the uranium dioxide $UO_2$. Our choice fell on this material for several reasons.
First, it is a simple system inducing a simplified numerical model:
- as an oxide it has a pure elastic behaviour (no plastic term need to be taken into account),
- oxidation occurs by oxygen incorporation in the solid (the chemical state of the sample can be described by the oxygen concentration only).

Second it undergoes the chemical fragmentation phenomenon in several different configurations, which makes possible its phenomenological analysis. $UO_2$ oxidation generates indeed different crystalline phases as a function of temperature and oxidation rate:
- at temperature higher than 400°C, $U_3O_8$ phase is mainly formed with a 36% swelling compared to $UO_2$.
- at temperature lower than 300°C, a layer of $U_4O_9$ cubic phase is first formed with 0.5% shrinking compared to $UO_2$, then a layer of $U_3O_7$ tetragonal phase appears on $U_4O_9$ layer. $U_3O_8$ formation occurs afterwards.
- at temperature around 300°C, $U_4O_9$ formation is not observed and only a $U_3O_7$ layer exists at low oxidation rate, before $U_3O_8$ formation.

In this paper, we will focus on the early $U_3O_7$ formation at 300°C, which can be seen as an $U_3O_7$ layer in epitaxy on a $UO_2$ substrate. The values used for the different internal parameters are given in *table 1*.

**Table 1**. Parameters values used in the simulations (ρ is the density).

|   | $D_1$ (cm$^3$/s) | $\eta_{ij}^1/\rho$ | T [°C] | E [GPa] | ν | Lattice parameters [Å] | | |
|---|---|---|---|---|---|---|---|---|
|   |   |   |   |   |   | a | b | c |
| Ref. | [9] | [11] |  | [10] |  | [7,8] | | |
| $UO_2$ | $0.0055 \exp\left(\frac{-26.3}{RT}\right)$ | $-1.248 \cdot 10^{-5}$ | 300 | 200 | 0,32 | 5,47 (cubic) | | |
| $U_3O_7$ |  |  |  |  |  | 5,40 (tetragonal) | | 5,49 |

Our results only refer to the oxygen dissolution for times lower than 1s. Indeed, the aim of the simulation is at the moment to study the parameters evolutions at the beginning of the oxidation process and to predict the crystalline phases liable to form. Moreover, we performed these numerical calculations to identify the differences induced on the oxygen diffusion due to different mechanical loadings, i.e. see the influence of different mechanical environments on this diffusion. So, a simplified 2 dimensional geometry was used: the sample introduced in the simulations is a parallelepiped one. It schematically represents a grain of material (*figure 2*). According to the classical diffusion processes, it corresponds to a substrate which could be oxidized. The hypothesis of semi-infinite sample imply $c_1(L_s) = 0$ and $\sigma(L_s)=0$ at the position $L_s$ far from external interface. Oxygen penetrates the material through the face $\Sigma$ and diffuses only on z direction.

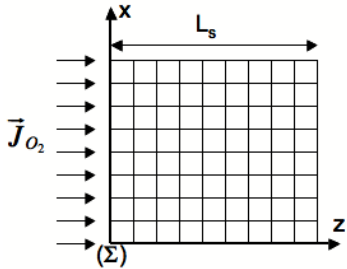

**Figure 2.** Simulated sample.

3.1. <u>No external stress applied</u>

It corresponds to the theoretical case where no interaction occurs between the grain and its environment. In this case we study both the material response to the dissolution of $O_2$ and the resulting internal stresses, i.e. the creation of stresses inside the material due to a purely chemical phenomenon. In the same way, such a simulation allows us to analyse the influence of the forced diffusion term in reference to the Fick's one. Let's break up the diffusion coefficient into three terms (equation 17). Different combinations are simulated: results are given in *figure 3*.

$$D_1' = D_1 + D_1 \frac{\eta_{ij}^1 c_1}{RT} \frac{\partial \sigma_{ij}}{\partial c_1} + D_1 \frac{\eta_{ij}^1}{RT} \sigma_{ij} = a + b + c \qquad (17)$$

*Figure 3* gives, as a function of the distance to the interface $\Sigma$:
- the evolution of oxygen concentration in the material,
- the evolution of the internal stress $\sigma_{xx}$.

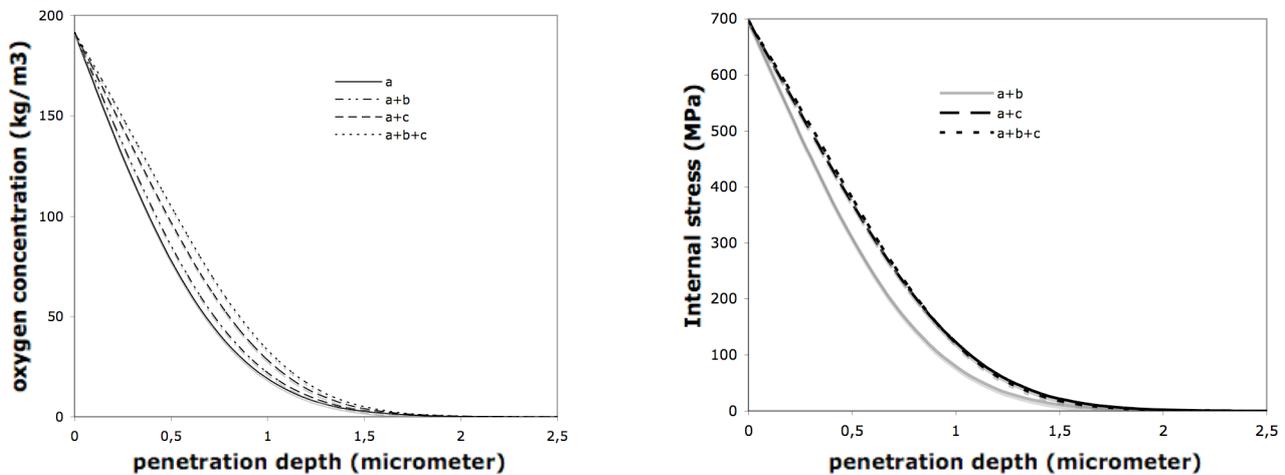

**Figure 3.** Oxygen concentration and internal stress $\sigma_{xx}$ evolutions versus penetration depth, for different expressions of the diffusion coefficient.

Concerning the dissolution curves, we notice an acceleration of the diffusion process when the forced diffusion terms are taken into account. However, the general shape of the diffusion curves stay unchanged. So, the stresses induced by the $O_2$ dissolution increase the diffusion coefficient values.

3.2. Simulation with an oxide layer

We want to determine the consequences of an interfacial deformation due to the oxide formation on the substrate's behaviour. Let's assume that the lattice adaptation between the substrate and its oxide generates deformations at the interface. Then, stresses appear, which can influence the oxygen diffusion in the material. In a first step, to simplify the calculation, we consider the different mechanical states created as a function of c axis orientation. We simulate then the presence of an oxide layer by considering two simple configurations (presented in *figure 4*):
- in the case (B), the oxide lattice parameter aligned to the sample surface is smaller than the substrate's one. To link the two lattices, it is necessary to expand the oxide layer. This expansion leads to compressive $\sigma_{xx}$ stress on the substrate surface.
- in the case (C), the oxide lattice parameter parallel to the sample surface is higher than the substrate's one. The linking of the two lattices induces a tensile $\sigma_{xx}$ stress on the substrate surface.

As comparison, case (A) gives the material response without oxide layer, i.e. without induced stress.

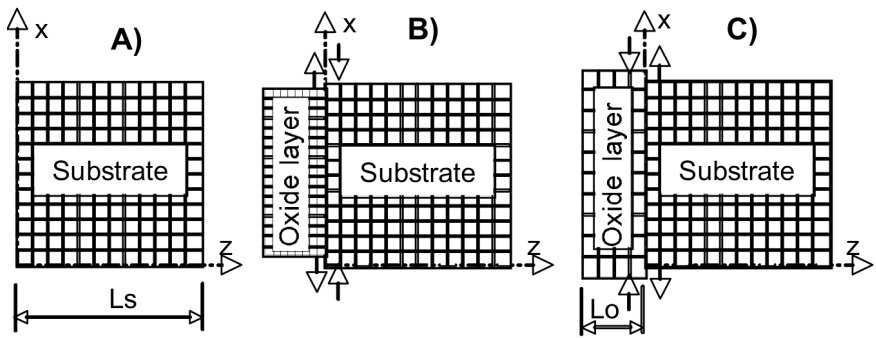

**Figure 4**. The different stress fields applied to the sample at internal interface ($L_0$=1µm).

3.3. Theoretical results

We give, according to the distance to the interface between the substrate and its oxide (called Σ):
- the evolution of oxygen concentration in the material,
- the evolution of the internal stress $\sigma_{xx}$.

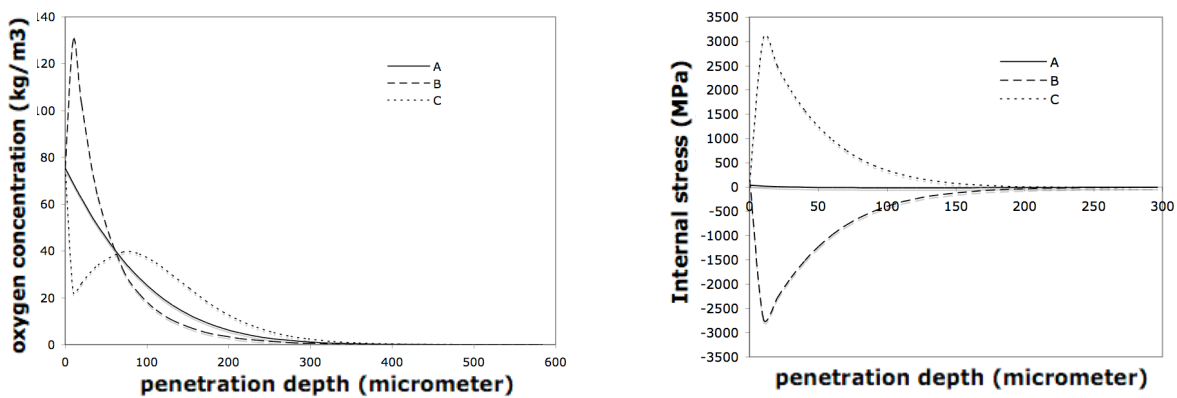

**Figure 5.** Oxygen concentration and internal stress $\sigma_{xx}$ evolutions vs. penetration depth into the substrate for three different imposed strain fields.

Results are given in *figure 5*. Even if it is still difficult to confirm experimentally the reached level of stresses calculated in these simulations, because the interface does not move in this approach, it is interesting to observe strong behaviour differences between the two simulated cases. If the formed oxide layer tends to distend the substrate's crystalline lattice (positive tensile stress, what corresponds to case C), we notice that the oxygen concentration slumps near the interface $\Sigma$ and increases when penetrating $UO_2$. This shows a slow down of the oxidation process due to strong oxygen dissolution in the volume. On the contrary, if the formed oxide layer compresses the crystalline lattice (negative stress, which corresponds to case B), the oxygen cumulates near the interface $\Sigma$. Then, in the case B, the oxide layer formation is made easier than in the case C, because the oxygen saturation is rapidly reached near the surface. Then, although the simulations represent an approximation of the real process, this case would correspond to the experimentally observed formation of $U_3O_7$ upon $UO_2$.

Concerning the case C, the very original behaviour it induces in $UO_2$ deserves to be analysed more precisely, in comparison with the behaviours observed experimentally in the formation of the different phases inside the oxide layer (micro-cracking).

4. CONCLUSION

A model is proposed to explain firstly the appearance of stresses at the substrate/oxide interface during the material oxidation, and secondly the influence of these stresses on the oxidation process evolution. For this occasion, a new expression of the diffusion coefficient, taking into account the stresses, is done. The simulations applied to the $UO_2/U_3O_7$ system show a great correlation between the lattice orientation and size of the oxide layer growing at the interface with the substrate and the induced stresses created, and consequently the possibility to form the oxide.

This first mechano-chemical approach of the diffusion mechanisms into a solid subjected to a strain field will be followed by a more detailed reconstruction of the material oxidation kinetic. This second step of our approach will be carried out soon, and based on a thermodynamic study of the solid/solid reactive interface displacement speed. This dynamic study will be completed by a static study of a multi-layer system ($UO_2/U_4O_9/U_3O_7$), which will allow us to evaluate the crystallographic compatibility of the lattices with regards to the reactional schemes. For this occasion, the strains generated by the linking of the different lattices will be studied thanks to the Bollmann's method.